\documentclass[aps,prd,twocolumn,preprintnumbers]{revtex4}

\usepackage{graphicx}
\usepackage{epsfig,rotating}

\usepackage[dvips]{color}
\definecolor{Black}{named}{Black}
\definecolor{Red}{named}{Red}

\def\nue{\nu_e}
\def\nuebar{\bar\nu_e}
\def\nux{\nu_x}

\def\theta{\vartheta}
\def\epsilon{\varepsilon} 
\newcommand{\be}{\begin{equation}}
\newcommand{\ee}{\end{equation}}
\newcommand{\ba}{\begin{eqnarray}}
\newcommand{\ea}{\end{eqnarray}}

\begin{document}

\preprint{MPP--2003--21}

\title{Supernova pointing with low- and high-energy neutrino 
detectors}

\author{R.~Tom\`as, D.~Semikoz, G.~G.~Raffelt, M.~Kachelrie\ss, 
  and A.~S.~Dighe}
 
\affiliation{Max-Planck-Institut f\"ur Physik
(Werner-Heisenberg-Institut),
F\"ohringer Ring 6, 80805 M\"unchen, Germany}


\begin{abstract}

A future galactic SN can be located several hours before the 
optical explosion through the MeV-neutrino burst, exploiting the
directionality of $\nu$-$e$-scattering in a water Cherenkov 
detector such as Super-Kamiokande.
We study the statistical efficiency of different methods for 
extracting the SN direction and identify a simple approach that is 
nearly optimal, yet independent of the exact SN neutrino spectra.  
We use this method to quantify the increase in the pointing accuracy 
by the addition of gadolinium to water, which tags neutrons from the 
inverse beta decay background.
We also study the dependence of the pointing accuracy on 
neutrino mixing scenarios and initial spectra.
We find that in the ``worst case'' scenario the pointing accuracy 
is $8^\circ$ at 95\% C.L.\ in the absence of tagging,
which improves to $3^\circ$ with a tagging efficiency of 95\%.
At a megaton detector, this accuracy can be as good as $0.6^\circ$.
A TeV-neutrino burst is also expected to be emitted contemporaneously
with the SN optical explosion, which may locate the SN to within 
a few tenths of a degree at a future km$^2$ high-energy neutrino 
telescope.
If the SN is not seen in the electromagnetic spectrum, locating it
in the sky through neutrinos is crucial for identifying the
Earth matter effects on SN neutrino oscillations.

\end{abstract}


\maketitle

\section{Introduction}

Observing a galactic supernova (SN) is the holy grail of low-energy
neutrino astronomy. The question ``how well can one locate the SN 
in the sky by the neutrinos alone?'' is important for two reasons.
Firstly, the MeV-neutrino burst precedes the optical explosion by 
several hours so that an early warning can be issued to the astronomical
community~\cite{Scholberg:1999tm,snews2}, 
specifying the direction to look for the explosion.
Secondly, in the absence of any SN observation in the electromagnetic 
spectrum, a reasonably accurate location in the sky is crucial for 
determining the neutrino Earth-crossing path to various detectors
since the Earth matter effects on SN neutrino oscillations may well 
hold the key to identifying the neutrino mass hierarchy
\cite{Dighe:1999bi,Lunardini:2001pb,Lunardini:2003eh,Takahashi:2001dc,%
Dighe2003,Dighe:2003jg}.

Nearly contemporaneously with the optical explosion an outburst of 
TeV neutrinos is expected due to pion production by protons accelerated 
in the SN shock~\cite{Waxman2001}.
This neutrino burst could produce of order 100 events in a future
km$^2$ high-energy neutrino telescope, allowing for a pointing
accuracy of a few tenths of a degree.  
Apart from the precise SN pointing, the detection of high-energy
neutrinos would be important as the first proof that SN remnants 
accelerate protons.

The optical signal from a SN, if observed, can give the 
most accurate determination of its position in the sky.
Apart from the observations at the optical telescopes,
the multi-GeV to TeV photons associated with the accelerated protons 
in the shock could be detected on the ground by air Cherenkov telescopes 
after the SN environment becomes transparent to high-energy photons. 
If a suitable x- or $\gamma$-ray satellite is in operation
at that time, the SN would be visible in these wavebands starting from
the optical explosion. A satellite like INTEGRAL could resolve the SN
with an angular resolution of 12~arc-minutes~\cite{integral}.

However, it is possible that the SN is not seen in the entire 
electromagnetic spectrum. 
This can be the case if it is optically obscured, no suitable x- or 
$\gamma$-ray satellite operates, and the air Cherenkov telescopes are 
blinded by daylight or the satellites and telescopes
simply do not look in the right direction at the right time.  It is
also possible that not every stellar collapse produces an explosion so
that only neutrinos and perhaps gravity waves can be observed. 
In such a scenario, the best way to locate a SN by its core-collapse 
neutrinos is through the directionality of $\nu e^- \to \nu e^-$ 
elastic scattering in a water Cherenkov detector 
such as Super-Kamiokande~\cite{Beacom:1998fj,sato}.  
Much less sensitive methods include the time-of-arrival 
triangulation with several detectors ~\cite{Beacom:1998fj,triang} or 
the systematic dislocation of neutrons in scintillation
detectors that measure $\bar\nu_e p\to n e^+$ and the subsequent
neutron capture~\cite{Apollonio:1999jg}.

The pointing accuracy of Super-Kamiokande or a future megaton
detector such as Hyper-Kamiokande or UNO is strongly degraded by the
inverse beta reactions $\bar\nu_e p\to n e^+$ that are nearly
isotropic and about 30--40 times more frequent than the directional
scattering events.  Recently it was proposed to add to the water a
small amount of gadolinium, an efficient neutron absorber, that would
allow one to detect the neutrons and thus to tag the inverse beta
reactions~\cite{vagins}.  
Evidently this would greatly improve the pointing:
a tagging efficiency of 90\% would double the pointing accuracy
\cite{vagins}.
At high tagging efficiency, however, the nearly isotropic
oxygen reaction $\nu_e+{}^{16}{\rm O}\to{\rm X}+e^-$
remains as the dominant background limiting the pointing accuracy.
In this paper we analyze the realistic pointing accuracy of a
water Cherenkov detector as a function of the neutron tagging
efficiency.

The directionality of the elastic
scattering reaction is primarily limited by the angular resolution of
the detector and to a lesser degree by the kinematical deviation of
the final-state positron direction from the initial neutrino.
Extracting information from ``directional data'' is a field in its own
right~\cite{FLE,Mardia}.  
An efficient method is the ``brute force'' maximum likelihood estimate
of the electron events, taking into account the angular
resolution function of the detector on top of a nearly isotropic
background. For a large number of events, the accuracy with this method 
in fact asymptotically approaches the minimum variance as given by the 
Rao-Cram\'er bound~\cite{Rao,cramer}.
However, for a small number of signal events, $N_s\alt 200$, we find
that a fraction of the information content of the data as
measured by the Fisher information~\cite{fisher-paper} cannot be
extracted by even the maximum likelihood method.
Using the Rao-Cram\'er bound
therefore overestimates the pointing accuracy of an experiment
for small $N_s$.
In this paper, we determine the realistic accuracy by using a 
concrete and nearly optimal estimation method.

Since the angular resolution depends on the
event energy, the likelihood method requires as input the functional
form of the neutrino energy spectra that are only poorly known.
It is difficult to systematically take into account the errors
introduced by a wrong choice of the fit function for the
likelihood method.
Therefore we look for ``parameter-free'' methods 
like harmonic analysis that use only 
the information contained in the data 
and exploit the symmetry of the physical situation.
We discuss the efficiency of two methods closely related to the 
harmonic analysis and find a simple iterative procedure making them 
nearly as efficient as the maximum likelihood approach.  
We use the most efficient method thus obtained for analyzing the
simulated events at a detector.
This makes our estimations realistic and even a bit conservative, 
since the existence of a better parameter-free method is not excluded.
We also study the dependence of the pointing accuracy on the neutrino
mixing parameters and the initial neutrino spectra, and use the
``worst case'' scenario in order to estimate the accuracy.

Finally, we briefly study the pointing accuracy in high-energy
neutrino telescopes that can pick up the TeV neutrino burst
expected around the time of the optical explosion.  While an accurate
pointing is easy for any of the existing and future neutrino
telescopes if the neutrinos are observed through the Earth, it is far
more difficult against the atmospheric muon background from
above. Even this would be possible for future km$^2$ detectors such as
IceCube or Nemo that could detect around $100$~SN events with TeV
energies within about one~hour.

We begin in Sec.~\ref{toy} with a discussion of the statistical 
methodology for extracting information from directional data 
using a toy model.  
In Sec.~\ref{simulation}
we study the realistic SN pointing accuracy of water Cherenkov
detectors as a function of the neutron tagging efficiency,
using realistic SN neutrino spectra.
In Sec.~\ref{tev} we turn to the pointing accuracy of
high-energy neutrino telescopes. 
Sec.~\ref{concl} is given over to conclusions.

\section{Analyzing directional data}
\label{toy}

\subsection{Pointing with maximum likelihood estimate}

\label{sec:lowerbounds}

As a first case we study  SN pointing with the $\nu$-$e$ elastic
scattering events in the absence of any other background, 
thus obtaining a bound on the realistic pointing accuracy.  
To this end we
use the toy model introduced in Ref.~\cite{Beacom:1998fj}, i.e.\ we
imagine a directional signal that is distributed as a two-dimensional
Gaussian on a sphere.  This choice is motivated by the observation
that the scatter of signal event directions is dominated by the
angular resolution of the detector, and by the assumption that the
angular resolution function is Gaussian. Later in Sec.~\ref{simulation} 
we consider a more realistic approximation to the angular detector response.

The angular width of the assumed Gaussian distribution is denoted by
$\delta_s$, where $s$ stands for ``signal.''  
As a further simplification we assume $\delta_s\ll\pi/2$,
allowing us to approximate the sphere by a plane.
Taking the signal to be centered at $\theta_0=0$, 
the probability
distribution function (pdf) of the signal events is
\begin{equation}
f_s(\theta,\phi) d\theta d\phi = \frac{1}{\cal C} 
\exp\left(-\frac{\theta^2}{2\delta_s^2}\right) d\mu
\end{equation}
with $d\mu= \sin \theta d\theta d\phi$. Here ${\cal C } \equiv \int
d\mu\exp[-\theta^2/(2\delta_s^2)]$ is a normalization constant taking the
value ${\cal C} = 2 \pi \delta_s^2$ for planar geometry.

In the case of Super-Kamiokande, around 300 elastic scattering events
constitute the directional signal. Assuming the mean electron energy
to be 11~MeV, a cone with opening angle $\ell_{68}\approx 25^\circ$
around the true direction contains 68\% of the reconstructed
directions~\cite{Nakahata:1998pz}. Solving
\begin{equation}\label{ds}
\int_{0}^{2\pi} d\phi \int_{0}^{(\ell_{68}\pi/180^\circ)} 
\! d\theta \, f_s(\theta,\phi) = 0.68 ~,
\end{equation}
we find $\delta_s \approx 17^\circ$.  
For $N_s~\gg~1$ signal events without background,
the Central Limit Theorem implies that the mean
reconstructed direction is within $\delta_s/\sqrt{N_s}$ of the true
direction for 68\% of all SN obervations.  This quantity, which is
$\approx 1^\circ$ for 300 events, gives the absolute lower bound on
the pointing accuracy in the absence of all backgrounds.  We note that
Ref.~\cite{Beacom:1998fj} uses $\delta_s \approx 25^\circ$ and hence
obtains $1.5^\circ$ for the pointing accuracy.  We further note that
our $\delta_s/\sqrt{N_s} \approx 1^\circ$ implies that 95\% of all SN
reconstructions lie within a circle of angular radius $2.4^\circ$
of the true direction.

The main degradation of the pointing accuracy is caused by the nearly
isotropic inverse beta decay background. The extent of this
degradation was addressed for the first time in
Ref.~\cite{Beacom:1998fj}.  The pdf on a sphere that represents $N_s$
signal events distributed like a Gaussian around the direction
$(\theta_0, \phi_0)$ as well as the $N_b$ isotropic background events
is
\begin{eqnarray}\label{eq:pdf}
&&f(\theta, \phi | \theta_0, \phi_0)\, d\theta\, d\phi\nonumber\\
&&=\,\frac{d\mu}{N_b+N_s} \left[ \frac{N_b}{4 \pi} +
\frac{N_s}{{\cal C}} \exp \left(-\frac{\ell^2}{2 \delta_s^2} \right)
\right] ~,
\label{full-pdf}
\end{eqnarray}
where 
\be
\ell \equiv \cos^{-1}[\cos \theta \cos\theta_0 + 
\sin \theta \sin \theta_0 \cos(\phi-\phi_0)]
\label{r-def}
\ee
is the angular distance between the direction of an incoming neutrino
and the experimentally measured direction of the Cherenkov cone. We
have introduced here the usual notation $f(x|x_0)$ for the pdf to
stress the dependence of $f$ on the data $x=(\phi,\theta)$ and
the parameters $x_0=(\phi_0,\theta_0)$.

The maximum likelihood estimate (MLE) method is the most efficient way
to extract information from statistical data.  For the pdf of
Eq.~(\ref{eq:pdf}) the likelihood function for $N$ events is
\be
L (\theta_0, \phi_0) \propto \prod_{\alpha=1}^N  
f(\theta^{(\alpha)},\phi^{(\alpha)} | \theta_0, \phi_0) ~,
\label{likelihood}
\ee
where $(\theta^{(\alpha)}, \phi^{(\alpha)})$ are the coordinates of the 
$\alpha^{\rm th}$  event. 

One commonly uses the Fisher information matrix~\cite{fisher-paper} to
estimate a lower bound on the uncertainty of the parameters extracted
by the MLE method.  In our case the Fisher matrix is defined as
\be
F_{ij} \equiv \left\langle \frac{\partial^2 \ln L(\theta_0, \phi_0)}
{\partial \Theta_i \partial \Theta_j} \right \rangle ~,
\label{fisher}
\ee
where $i,j=1,2$, $\Theta_1 \equiv \theta_0$, $\Theta_2\equiv\phi_0$
and $\langle\ldots\rangle$ denotes an average with respect to
$fd\theta d\phi$.  Since the off-diagonal elements of this matrix
vanish, the error in the measurement of the two angles is simply given
by
\be
\Delta \Theta_i = \sqrt{1/F_{ii}} ~.
\label{fisher-error}
\ee
This lower bound on the pointing error is also known as the
Rao-Cram\'er bound~\cite{Rao,cramer}.

For the sake of definiteness, we have chosen the SN direction 
to lie in the equatorial plane so that $\Delta\theta_0=\Delta\phi_0$, 
and define the pointing error $\Delta\theta$ as 
\be
\Delta\theta \equiv \sqrt{ \frac{1}{N-1} 
\sum{(\bar\theta- \langle \bar\theta \rangle)^2}}~,
\ee
where $\bar\theta$ is the estimate for $\theta_0$ from a given method
and $N$ is the number of simulations used.
The Rao-Cram\'er bound corresponds to the inequality 
\be
(\Delta\theta)^2 \geq (\Delta\theta)^2_{\rm Fisher} \equiv 1/F~,
\ee
where $F$ may be $F_{11}$ or $F_{22}$.

The accuracy of the MLE method approaches asymptotically the
Rao-Cram\'er bound for a sufficiently large number of
events~\cite{stat-book}. For a finite number of data points, the MLE
saturates this bound only 
if
the pdf used in the MLE can be written in the product form
\be
 f(x | x_0) = g(x_0)h(x) \exp[A(x_0) B(x)] ~.
\ee
If the pdf cannot be written in this form, the efficiency of the MLE
is significantly smaller than unity at low values of $N_s$.

\begin{figure}[ht]
\includegraphics[height=0.48\textwidth,clip=true,angle=0]{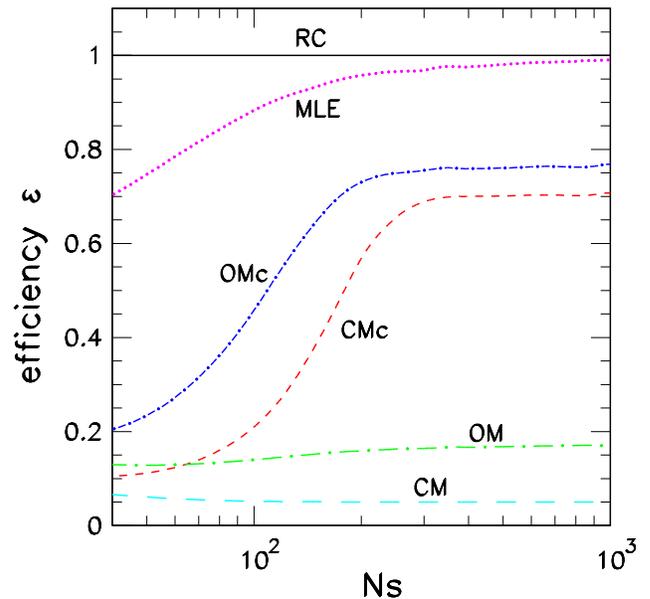}
\caption[...]{Efficiencies of different estimation methods
described in the text for $N_b/N_s = 30$.
RC corresponds to the Rao-Cram\'er minimum variance.
 \label{fig:eff}}
\end{figure}

In Fig.~\ref{fig:eff} we show the efficiency $\epsilon$ of the MLE 
in ``Fisher units'',
\be
\epsilon  \equiv (\Delta\theta)^2_{\rm Fisher} /
(\Delta\theta)^2 ~,
\ee
as a function of the number $N_s$ of signal events while keeping the 
background-to-signal ratio $N_b/N_s$  fixed at~30,
which is the expected ratio of the inverse beta decay events to the
elastic scattering events in a water Cherenkov detector.
Though the MLE efficiency tends asymptotically 
to the Rao-Cram\'er bound for large values of $N_s$,
this bound overestimates the MLE pointing accuracy by 
$\sim 10$\% for $N_s\alt 200$.

We stress that the Rao-Cram\'er bound as an estimate of the MLE
pointing efficiency is useful only in the limit where the spherical
nature of our problem can be approximated by a planar geometry, i.e.\
when the angular resolution of the detector is much better than
$90^\circ$. This condition is satisfied in our case.  For
the general case of a pdf defined on a sphere the MLE still provides
an optimal method to determine the pointing accuracy, but the Fisher
information matrix no longer provides a direct asymptotic bound
on the pointing accuracy.
We are not aware of a generalized version of the Rao-Cram\'er bound that
would relate directly to the variance of the pointing estimate in the 
case of a truly spherical problem. 

\subsection{Efficiencies of parameter-free methods}
\label{cg-mi}

The MLE is an optimal method to extract information from experimental
data if the probability distribution function is
known. This is not the case in our situation, 
where the exact forms of the neutrino spectra are needed
and these are only poorly known. 
It is therefore
worthwhile to look for other methods which may be less efficient, but
which do not depend on the exact form of the pdf. 
In our case of SN
pointing we wish to consider methods that do not depend on prior
knowledge of the exact neutrino energy spectra.

Let us consider two pointing methods that exploit the symmetries of
our physical situation, but are independent of the exact details of
the pdf. If the efficiency of such a method turns out to be comparable
to the maximum likelihood method for the toy model of the previous
section, then that method may be expected to be efficient also in the
realistic case where the exact pdf is not known and the MLE method
cannot be employed.

One obvious approach is the ``center of mass'' (CM) method. The center
of mass $\bf{S}$ of the events,
\be
 S_i \equiv \sum_{\alpha=1}^N m^{(\alpha)} x_i^{(\alpha)}
 \,, \qquad i=1,2,3~,
\label{cm}
\ee
is taken to be the estimator of the true center of the distribution,
where $\bf{x} \equiv
(\sin\theta\cos\phi,\sin\theta\sin\phi,\cos\theta)$.  For an ideal
detector the event weights $m^{(\alpha)}$ are equal and can be set to
one.  More realistically, the detection probability $p$ is a function
of the detection angles $\theta,\phi$ and the weight is
$m^{(\alpha)}=1/p(\theta^{(\alpha)},\phi^{(\alpha)})$.

A second approach is the ``orientation matrix'' (OM)
method~\cite{FLE}. Here, the major principal axis of the orientation
matrix
\be 
 T_{ij}\equiv
\sum_{\alpha=1}^N m^{(\alpha)} x^{(\alpha)}_i x^{(\alpha)}_j
\label{om}
\ee
is taken to be the estimator. 

These two methods are equivalent to a harmonic analysis,
\be  
\label{ha}
a_{lm}= \sum_{\alpha=1}^N Y_{lm}(\theta^{(\alpha)},\phi^{(\alpha)}) ~,
\ee
restricted to the first moment for CM, and up to the second moment for
OM.  This equivalence can be seen either by a direct evaluation of
Eq.~(\ref{ha}) or by identifying $S_i$ with a dipole moment and
relating $T_{ij}$ to a quadrupole moment $Q_{ij}$. Since
$3T_{ij}=Q_{ij}+ N\delta_{ij}$, the direction of the major principal
axis is identical for both ellipsoids. Note that the orientation
matrix is a reducible tensor and therefore contains information from
the first as well as the second moment.

\begin{figure}[ht]
\includegraphics[height=0.48\textwidth,clip=true,angle=0]{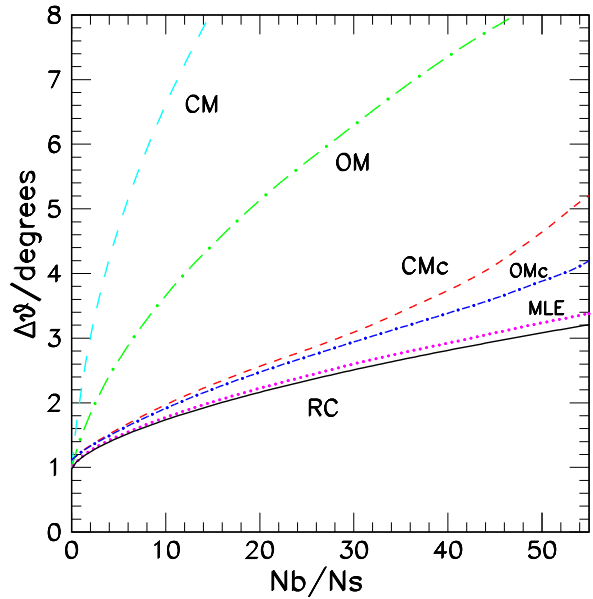}
\caption[...]{Pointing error $\Delta \theta$ for different estimation 
methods for  $N_s = 300$. RC corresponds to the minimum variance as
given by the Rao-Cram\'er bound.
\label{fig:deltatheta}}
\end{figure}

Neither of these methods requires any prior knowledge of the neutrino
spectra or cross sections. However, they involve some loss of
information and hence will give larger pointing errors than the
MLE. In order to quantify the efficiency of these methods we generate
a data sample according to the pdf of Eq.~(\ref{full-pdf}) and show
the respective pointing errors in Fig.~\ref{fig:deltatheta}.  We keep
the number of signal events fixed at $N_s=300$, and show the pointing
error $\Delta\theta$ as a function of $N_b/N_s$.  
Note that in the absence of neutron tagging this ratio is expected
to be around 30--40.

Figure \ref{fig:deltatheta} shows that the error of MLE is
almost the same as the Rao-Cram\'er (RC) bound.  However, the errors
of CM and OM are much larger.  One may also notice that OM is more
accurate than CM.  This difference may be attributed to the fact that
whereas CM tries to exploit the spherical symmetry of the background,
OM exploits the cylindrical symmetry of the background about the
arrival direction, which is broken much more weakly by statistical
fluctuations in the background.  Moreover, in terms of a harmonic
analysis, OM involves information from both $l=1$ and 2 while CM
involves only $l=1$.

In order to increase the efficiency of CM and OM, we use the physical
input that the signal is concentrated within a small region around the
peak. Cutting off the events beyond a certain angular radius would
then increase the signal to background ratio and the above methods may
be applied iteratively to this new data.  
This procedure converges
quickly and gives a much better estimate of the incoming neutrino
direction. The optimal value of the angular cut has a very weak
dependence on the number of events and the background-to-signal ratio.
It depends mainly on the value of $\delta_s$ and is found to lie
between $2\delta_s$ and $3 \delta_s$. 
Within this range, the efficiency depends only weakly on the exact 
value of the angular cut. 
We tried both a sharp cutoff and a Gaussian weight function; both
choices give practically identical results.

The optimal value of the cut also increases slowly with decreasing
background-to-signal ratio, and in the limit of zero background,
the method without cut is clearly more efficient than the 
method with cut since the latter now cuts off signal but no background.
However, the variation due to changing the cut is of the order 
of only a few percent.
Therefore, we keep the value of the cut
to be constant at $40^\circ$ for the analysis in this section.
Our results for the pointing accuracy will therefore be
somewhat conservative.

We denote the CM and OM methods with this cutting procedure by CMc and
OMc, respectively.  Figure \ref{fig:deltatheta} shows how the 
pointing error decreases drastically with the cutting procedure. 
It may also be observed that the accuracy of CMc stays close to that of 
MLE for low values of the background-to-signal ratio, while the accuracy of 
OMc is close to MLE in the entire $N_b/N_s$ range.

The efficiencies of all methods depend on both $N_s$ and $N_b/N_s$.
In Fig.~\ref{fig:eff} we show the efficiency $\epsilon$ for different
methods as a function of $N_s$, keeping $N_b/N_s$ fixed at 30.  
For a large number of signal events, $N_s\agt 300$, all methods tend
to their asymptotic efficiencies $\epsilon_\infty$.  The OMc method is
close to its asymptotic efficiency $\epsilon_\infty\approx 0.77$
already for $N_s\approx 200$ whereas CMc needs $N_s\approx 300$ events
to reach $\epsilon_\infty\approx 0.71$.
Since the OMc method turns out to be more efficient that CMc
in all the parameter ranges, henceforth we continue using only the
OMc method for further estimations.

\begin{figure}[ht]
\includegraphics[height=0.48\textwidth,clip=true,angle=0]{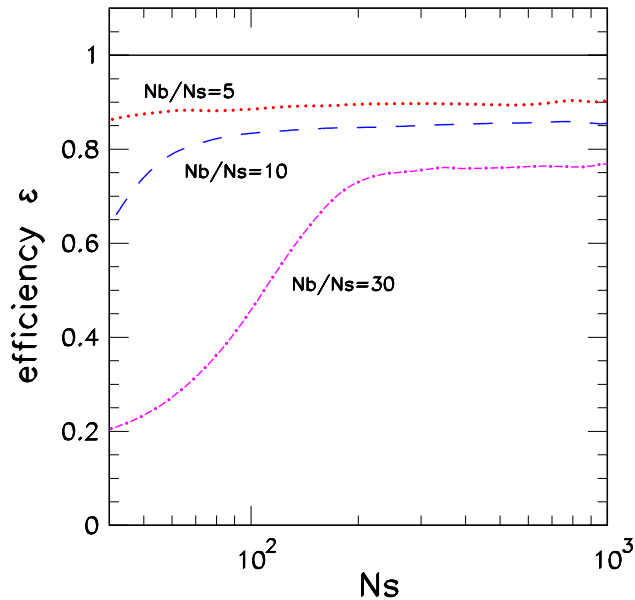}
\caption[...]{Efficiency of OMc with a $40^\circ$ angular cut for 
different values of $N_b/N_s$. \label{fig:moreeff}}
\end{figure}

With neutron tagging, the value of $N_b/N_s$ decreases, and that 
increases the efficiency of the OMc method. In Fig.~\ref{fig:moreeff},
we show the efficiency of OMc at different values of $N_b/N_s$.
At $N_b/N_s = 10$, which corresponds to the tagging efficiency
$\epsilon_{\rm tag} \approx 67\%$, the asymptotic efficiency of OMc already
increases to 0.85, and at $N_b/N_s = 5$, corresponding to 
$\epsilon_{\rm tag} \approx 83\%$, it reaches 0.90. 
Moreover, with decreasing $N_b/N_s$ the asymptotic value is reached at 
lower and lower number of signal events. 
For higher tagging efficiencies, the optimal value of the angular cut
increases. In fact, as noted before,
in the limit of no background the OM method without
the cut is more optimal. However this limit is physically not 
reached due to the presence of oxygen  events.

The OMc method thus sacrifices less than 25\%, and at higher
tagging efficiency, even less than 15\%, of the
pointing accuracy of the MLE method.
On the other hand, it  has the great advantage of
being independent of the detailed neutrino spectra and cross sections.
Therefore, this method can be extremely useful for a fast analysis of
the SN signal. After all, an early warning would depend on a quick and
simple data analysis while later one can certainly optimize by fitting
detailed energy spectra to the observed signal.

We stress that our preference for a parameter-free
method over MLE in this analysis is strongly influenced by the current 
status of our knowledge regarding the pdf of the angular distribution. 
It may indeed be possible to use MLE and include all the systematic 
uncertainties, perhaps giving a better estimate for the pointing
accuracy.
However, faced with a tradeoff between model independence and 
higher efficiency, we give more weight to the former.
If in future we understand the primary spectra much better than we 
do now, this preference may change.

\section{Supernova pointing accuracy of water Cherenkov detectors}
\label{simulation}

We now apply our method to a more realistic
representation of the SN signal in a water Cherenkov detector.
We shall limit our analysis of the pointing
accuracy to our best parameter-free method, i.e.\ the Orientation
Matrix method with an angular cut (OMc).

In order to determine the pointing accuracy numerically we simulate a
large ensemble of SN signals in a water Cherenkov detector, assuming
different efficiencies for neutron tagging.  To this end we assume
that the SN is at a distance $D=10$~kpc and releases the neutron-star
binding energy $E_b=3\times 10^{53}$~erg in the form of
neutrinos. Details of the assumed neutrino spectra and fluxes are
given in Appendix~\ref{nu}.  
The spread in the predicted neutrino spectra has been taken care of
by using two models, a model from the Garching group (model G)
\cite{garching-model} and a model from the Livermore group
(model L) \cite{livermore-model} as described in the same appendix.
We take into account the effects of neutrino 
flavor conversions by considering the three mixing scenarios, 
(a) normal mass hierarchy and $\sin^2\Theta_{13} \agt 10^{-3}$,
(b) inverted  mass hierarchy and $\sin^2\Theta_{13} \agt 10^{-3}$, and
(c) any mass hierarchy and $\sin^2\Theta_{13} \alt 10^{-3}$.
The six combinations of the models and neutrino mixing scenarios
are represented by G-a, G-b, G-c, L-a, L-b, L-c.
We use $\sin^2(2\Theta_\odot) = 0.9$ for the solar neutrino
mixing angle.

As reaction channels we use elastic scattering on electrons
$\nu e^-\to\nu e^-$, inverse beta decay $\bar\nu_e p\to n e^+$, and
the charged-current reaction $\nu_e+{}^{16}{\rm O}\to X+e^-$, while
neglecting the other, subdominant reactions on oxygen.  The cross
sections for these reactions are summarized in Appendix~\ref{cross}.
The oxygen reaction is included because it provides the dominant
background for the directional electron scattering reaction in a
detector configuration with neutron tagging where the inverse beta
reaction can be rejected.

For the detector we assume perfect efficiency above an
``analysis threshold'' of 7~MeV, and a vanishing efficiency 
below this energy.
The actual detector threshold may be as low as 5~MeV.
Though lowering the threshold increases the ratio of elastic scattering 
events and the inverse beta events, it also introduces a background 
from the neutral-current excitations of oxygen 
(see Appendix~\ref{cross}). 
In order to avoid additional uncertainties from the cross section
of these oxygen reactions, we use the higher analysis threshold.
We have checked that the net improvement by lowering the
threshold to 5 MeV is less than 10\% in all cases.

We assume a fiducial detector mass of 32 kiloton of water.
Using the neutrino spectra and mixing parameters from the six cases
mentioned above, we obtain 250--300 electron scattering events,
7000--11500 inverse beta decays, and 150--800 oxygen events. 
The ranges correspond to the variation due to the six different 
combinations of  neutrino mixing scenarios and models for the
initial spectra.

\begin{figure}[ht]
\includegraphics[height=0.48\textwidth,clip=true,angle=270]{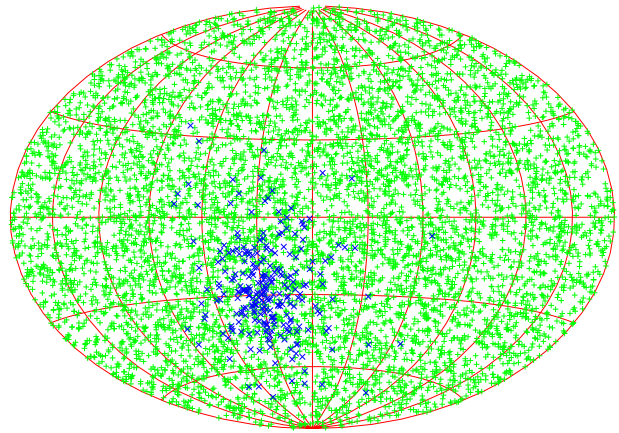}
\caption[...]{
Angular distribution of $\bar\nu_ep\to ne^+$ events (green) and
elastic scattering events $\nu e^-\to\nu e^-$ (blue) 
of one simulated SN.
\label{map}}
\end{figure}

The procedure of event generation is described in 
Appendix~\ref{ang-res}. The angular resolution function of the 
Super-Kamiokande detector does not follow a Gaussian distribution,
rather it is close to a Landau distribution that we use for our
simulation.
In Fig.~\ref{map}, the angular distribution of $\bar\nu_ep\to ne^+$
events (green) and elastic scattering events $\nu e^-\to\nu e^-$
(blue) of one of the simulated SNe are shown in Hammer-Aitoff
projection, which is an area preserving map from a sphere to a plane.

The position of the SN is estimated with the OMc method.
As explained in Sec.~\ref{toy}, the optimal value of the angular cut
depends on the neutron tagging efficiency as well as the 
neutrino spectra.
We use a sharp cutoff with $30^\circ$ opening angle for the OMc, 
which may not be optimal, but is observed to be close to optimal 
in almost the whole parameter range. 
For low values of $\epsilon_{\rm tag}$, the value of the cut should be 
lowered whereas for large values of $\epsilon_{\rm tag}$ it should be
increased by about $10^\circ$. The optimal cut depends also on the 
details of the detector properties and neutrino spectra.

\begin{figure}[h]
\epsfig{file=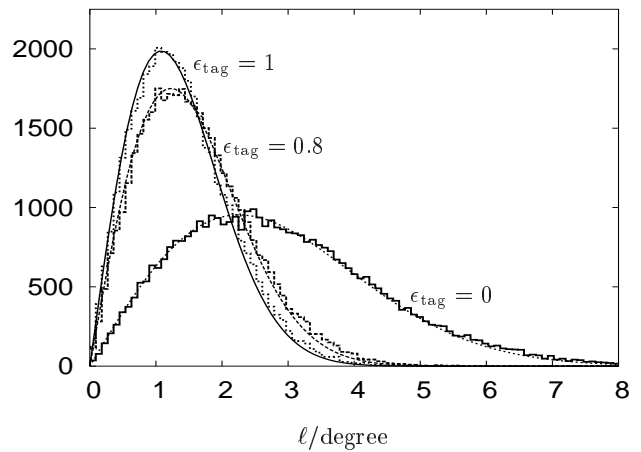,height=6.5cm} 
\caption{
Histogram of the angular distance $\ell$ of the estimated SN direction
to the true one for $40000$ simulated SNe with neutrino parameters
corresponding to G-a and neutron tagging efficiencies 
$\epsilon_{\rm tag}=0, 0.8$ and 1. The fits using the distribution
in Eq.~(\ref{fl}) are also shown.
\label{dist}}
\end{figure}

A histogram of the angular distances between the true and the estimated
SN position found in $40000$ simulated SNe for different neutron tagging
efficiencies for the case G-a is shown in Fig.~\ref{dist}. 
The histogram fits well the distribution 
\be
 f(\ell) d\ell = \frac{1}{\delta^2} 
\exp\left(-\frac{\ell^2}{2\delta^2}\right) \ell d\ell ~,
\label{fl}
\ee
where $\ell$ is the angle between the actual and the estimated 
SN direction, and $\delta$ is a fit parameter.

\begin{figure}[h]
\epsfig{file=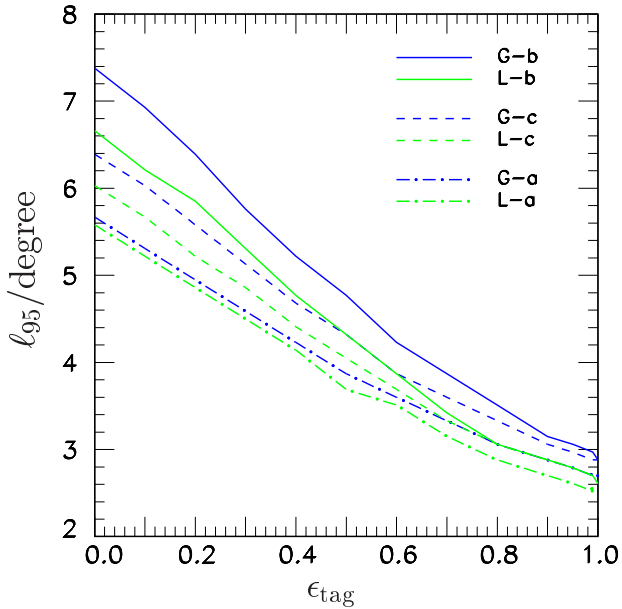,height=9.0cm}
\caption{
The pointing accuracy $\ell_{95}$ as a function of the neutron
tagging efficiency $\epsilon_{\rm tag}$ for six cases corresponding to
three neutrino mixing scenarios and two models for the initial
neutrino spectra. 
\label{fig:abc}}
\end{figure}

Defining the opening angle $\ell_\alpha$ for a given confidence level $\alpha$ 
as the value of $\ell$ for which the SN direction estimated by a fraction 
$\alpha$ of all the experiments is contained within a cone of 
opening angle $\ell$, we show in Fig.~\ref{fig:abc} the opening angle 
for 95\% C. L. for the six cases of neutrino parameters.
Clearly, the pointing accuracy depends weakly on the neutrino mixing 
scenario as well as the initial neutrino spectra. 
Some salient features of this dependence may be understood 
qualitatively as follows.

The signal events are dominated by $\nue$. Indeed, nearly half
of the elastic scattering events are due to $\nue$, whereas the remaining
half are due to the other five neutrino species. 
The cross section of electron scattering events increases with energy.
Therefore, the more energetic the $\nue$ arriving at the detector, 
the larger the number of signal events and the better the 
pointing accuracy.
Though the initial average $\nue$ energies are equal in the 
models G and L, the model L gives a much larger average energy for the 
initial $\nux$ spectrum.
The $\nue$-$\nux$ mixing then tends to give more energetic $\nue$ in the 
model L. As a result, for each mixing scenario, the  model L
predicts a better pointing accuracy than the model G.

Within a model, the pointing accuracy is governed by the 
background-to-signal ratio. 
Since the cross section of the dominating background $\bar{\nu}_e p$
reaction increases with energy, more $\nuebar$-$\nux$ mixing 
tends to give more energetic $\nuebar$ and hence more background 
and less pointing accuracy.
The ratio of  $\nuebar$-$\nux$ mixing to $\nue$-$\nux$ mixing 
within a model is the smallest for the mixing scenario (a) and 
the largest for the scenario (b). 
Therefore, within a model the scenario (a) always gives the best 
pointing  accuracy and the scenario (b) always gives the worst. 
Note that in the limit $\epsilon_{\rm tag}=1.0$ when all
the $\bar{\nu}_e p$ background is eliminated, the scenarios 
(b) and (c) give identical pointing accuracies since the final 
$\nue$ spectra in these two schemes are identical.

If neutrino experiments or supernova simulations have already identified the 
actual scenario, we could use the confidence limits given by that 
particular scenario. Indeed, the information for identifying the
mixing scenario may already be contained in the 
observed neutrino spectra themselves which may be extracted by 
further data analysis 
\cite{Dighe:1999bi,Lunardini:2001pb,Lunardini:2003eh,Takahashi:2001dc,%
Dighe2003,Dighe:2003jg}. 
However, in the absence of the immediate availability of this information,
one has to select the least efficient scenario, G-b, in order to
obtain the most conservative limits. This is the scenario with the least 
$\nue$-$\nux$ mixing, the largest $\nuebar$-$\nux$ mixing and the lowest 
initial average energy for $\nux$.

Since the pointing accuracy is worse when the component of initial
$\nux$ flux in the final $\nue$ flux is smaller, the ``worst case''
scenario is expected to be when the $\nue$ survival probability $p$ 
(see Appendix~\ref{nu}) takes its highest possible value, 
which is nearly 0.45 at 3$\sigma$ \cite{fit-paper} , 
corresponding to $\sin^2(2\Theta_\odot) \approx 0.99$. 
In Table~\ref{t1} we give the values of the opening 
angle $\ell$ for this ``worst case'' scenario for various confidence 
levels $\alpha$ and tagging efficiencies.
We also give the values of the fit for $\delta$ in Eq.~(\ref{fl}), from
which the numbers for any confidence level can be read off.
For $\epsilon_{\rm tag}=0$, at 95\% C.L.\ the pointing accuracy is 
$7.8^\circ$, which improves to $3.6^\circ$ for $\epsilon_{\rm tag} =80\%$ 
and $3^\circ$ for $\epsilon_{\rm tag} = 1$.
This is nearly a factor of 3 improvement in the pointing angle, 
which corresponds to almost an order of magnitude improvement in the area 
of the sky in which the SN is located.

\begin{table}
\caption{\label{t1} 
Opening angle $\ell_\alpha$ of the cone with $\alpha$ confidence level to
contain the true SN direction for different tagging efficiencies 
with the ``worst case'' scenario.
The bottom row gives the width $\delta$ of the Gaussian distribution
$f(\ell)$. The $\star$ column gives the pointing accuracy in the limit 
when all the background, including the oxygen events, is weeded out.}
\begin{ruledtabular}
\begin{tabular}{cccccccc}
& \multicolumn{6}{c}{\large $\epsilon_{\rm tag}$} &  \\
 & 0 & 0.5 & 0.8 & 0.9 & 0.95 & 1.0 & $\star$ \\
[0.5ex]\hline
$\alpha=0.68$ & 4.7 & 3.0 & 2.4 & 2.0 & 1.9 & 1.8 & 1.7 \\  
$\alpha=0.90$ & 6.8 & 4.2 & 3.1 & 2.9 & 2.7 & 2.6 & 2.5 \\  
$\alpha=0.95$ & 7.8 & 5.0 & 3.6 & 3.2 & 3.2 & 3.0 & 2.9 \\  
$\alpha=0.99$ & 10.0 & 6.1 & 4.5 & 4.1 & 3.9 & 3.7 & 3.6 \\  \hline
$\delta$ & 3.0 & 1.9 & 1.4 & 1.3 & 1.2 & 1.2 & 1.1 \\  
\end{tabular}
\end{ruledtabular}
\end{table}

The last column (marked by $\star$) in the table shows the
pointing accuracy in the limit of no background, i.e.\ the case
where all the inverse beta decay as well as the oxygen events are
weeded out. This gives the intrinsic limiting accuracy due to
the angle of electron scattering, the angular resolution of the
detector and the efficiency of our OMc algorithm. For 95\% C.L.\,
the table gives a pointing accuracy of $2.9^\circ$. This
may be compared with the $2.4^\circ$ that was estimated 
in Sec.~\ref{sec:lowerbounds} for the toy model 
when there was no background.
The degradation in the pointing accuracy may be attributed to the
loss of information in the OMc method, the 10\% smaller number of
events in the SN simulation, and the difference between the 
actual angular distribution and the Gaussian as taken in the toy model.

For a SN at 10 kpc, in the worst case scenario we get
nearly 10600 events, out of which the electron 
scattering signal is $N_s \approx 270$.
We are then already at or just below the asymptotic 
limit for the efficiency of the OMc method. 
For a larger detector like Hyper-Kamiokande, the desired accuracy 
can be calculated simply by rescaling according to the number of signal 
events. 
For a detector with 25 times the fiducial volume of Super-Kamiokande,
the pointing accuracy is then expected to be $2^\circ$ without gadolinium 
and $0.6^\circ$ with $\epsilon_{\rm tag} > 90\%$.
If the total number of events is much smaller than 10000, for
small $\epsilon_{\rm tag}$ the efficiency of OMc will be smaller
and this factor needs to be taken into account for calculating
the real pointing accuracy. 
For $\epsilon_{\rm tag}> 0.8$, the scaling with number of events
should even work for a very small number of events, as can be
seen from Fig.~\ref{fig:moreeff}.

Note that all the model dependences discussed here, though clearly
observable, are only around 10\% (see Fig.~\ref{fig:abc}).
This indicates that the pointing accuracy estimates are quite
model independent, and hence robust.

\section{High-energy neutrino telescopes}
\label{tev}

\subsection{High-energy supernova neutrinos}

Turning to the putative TeV neutrino burst associated with a SN
explosion we note that the shock wave may well accelerate protons to
energies up to $10^{16}$~eV.  This idea is supported by the fact that
the cosmic rays below the knee, $E < 10^{16}$~eV, contain a total
amount of energy comparable to that injected by all galactic SNe.  The
possibility of detecting high-energy neutrinos from a galactic SN has
been discussed in the literature.  In particular, it was shown that
during the first year after the explosion, the SN shock wave will
produce a large flux of neutrinos with energies above 100~GeV,
inducing more than $10^3$ muons in a km$^2$
detector~\cite{BerezinskyPtuskin}. More recently it was suggested that
the high-energy neutrino signal would arrive just 12~hours after the
SN explosion and would last for about one hour, giving  about
100~muon events with $E > 1$~TeV in a km$^2$
detector~\cite{Waxman2001}.

Of course, the number of expected events strongly depends on unknown
parameters, in particular on the total energy emitted in the form of
pions $E_\pi^{\rm tot}$ at a given time and the maximum energy of the
emitted neutrinos $E_{\rm max}$. The neutrino spectrum depends on that
of the accelerated protons. However, higher-energy neutrinos have a
greater chance of being detected so that we are not interested in the
exact spectral shape once it is a typical power-law for shock
acceleration, $dN_\nu/dE \propto E^{-\alpha}$ with $\alpha\leq2$.  If
the spectrum is softer, $\alpha>2$, detecting the neutrinos is more
difficult. Note that the spectrum of protons after shock acceleration
can be dominated by high-energy particles in some
cases~\cite{Derishev2003}, or even can be monochromatic if protons are
accelerated in a potential gap. We will concentrate on the
``standard'' case of a $E^{-2}$ neutrino spectrum, although our
results for other cases will be similar. Following the calculation of
Ref.~\cite{Waxman2001} and assuming 
$E_{\rm max}=1$~TeV we expect around 50~muon events in a km$^2$
detector during 1~hour at a time about 12~hours after the SN
explosion. For a larger maximum energy,
$E_{\rm max}=1000$~TeV, the number of muons increases to~200.

\subsection{Signal from below}

In order to discuss the expected signal in different neutrino
telescopes~\cite{Spiering2002,Halzen2002} we begin with the case where
the SN happens in a part of the sky that a given neutrino telescope
sees through the Earth. 
Future km$^2$ detectors like IceCube at the
South Pole~\cite{ICECUBE}, or northern projects like the
Gigaton Water Detector at Baikal~\cite{GVD} and Nemo in the
Mediterranean~\cite{NEMO} can detect the high energy neutrinos.
For each event the angular resolution is around one degree.  In this
case the pointing to the SN can be resolved with an accuracy of about
$1^\circ/\sqrt{50}\approx 8'$~(arc-minutes). However, this purely
statistical error does not include possible systematic effects. Most
important is the limited knowledge of the alignment of these
detectors. Therefore, the pointing accuracy of a km$^2$ detector is
probably larger and around a few tenths of a degree.

Even the existing smaller detectors can see a significant signal.  The
northern sky is under control of AMANDA-II with an effective area of
0.1~km$^2$ and angular resolution of $2^\circ$ at TeV energies. 
AMANDA-II will then detect 5 (20) events for $E_{\rm max}=1$~TeV (1000~TeV)
and thus will be able to resolve the SN direction to better than
$1^\circ$. 
After their completion, the northern projects ANTARES~\cite{antares}
and NESTOR~\cite{nestor} will be comparable to AMANDA-II.

\subsection{Signal from above}

If the high-energy SN neutrinos arrive ``from above,'' they are masked
by the large background of atmospheric muons. For IceCube 
this background is about $5.2\times10^{10}~{\rm year}^{-1}$ 
from the upper hemisphere at the ``trigger'' level \cite{ice-trigger}.
This corresponds to nearly $300~\rm hour^{-1}~degree^{-2}$. 
If we note that the angular
resolution is about $1^\circ$, the expected signal of 100~events
will be much larger than the background fluctuations in one pixel of
the sky. Morever, the expected SN neutrinos will have multi-TeV
energies so that energy cuts will reduce the background. Also, the
significance of the signal will be enhanced by the angular prior 
defined by the low-energy signal in Super-Kamiokande.

For AMANDA-II size detectors both background and signal are about
10~times smaller. Moreover, the angular resolution is only about
$2^\circ$.  Therefore, the expected signal in one pixel of the sky
will be comparable to the background fluctuations.  Energy cuts may
allow one to detect the SN signal from the ``bad'' side of the sky.

\section{Summary and Conclusions}
\label{concl}

The MeV neutrinos from the cooling phase of a SN will arrive at
the Earth several hours before the optical explosion. These
neutrinos will not only give an early warning of the advent of
a SN explosion, but they can also be used to determine the
location of the SN in the sky, so that the optical telescopes
may concentrate on a small area for the observation.

In a water Cherenkov detector like Super-Kamiokande, the $\nu$-$e$ 
scattering events are forward peaked and thus can be used for
the pointing. The main background comes from the inverse beta decay
reactions $\nuebar p \to n e^+$, which is nearly isotropic and
has a strength more than 30 times that of the electron scattering 
``signal'' reaction. 
The reactions of the neutrinos with oxygen also contribute to
the nearly isotropic background.

The authors of Ref.~\cite{Beacom:1998fj} have estimated the 
pointing accuracy  using the Rao-Cram\'er bound. 
This may overestimate the accuracy by $\sim 10$\%
since even the most efficient method, the maximum likelihood 
estimate, can reach the minimum variance bound only for 
a large number of signal events. 
More importantly, the maximum likelihood method needs as an input 
the exact form of the fit function, which is not available due to 
our currently poor knowledge of the neutrino spectra. 
Taking into account the errors due to a wrong choice of the
fit function is difficult. Therefore we choose to calculate the 
pointing accuracy using a concrete and simple estimation method 
that is independent of the form of the fit function.

We explore some parameter-free methods that only use the data
and exploit the symmetries inherent in the physical situation, 
and therefore
give a model independent estimation of the pointing accuracy
while sacrificing some information from the data.
We perform a statistical analysis of these methods using a toy model,
which has a Gaussian signal on top of an isotropic background.
We find that a method that uses the ``orientation matrix''
with an appropriate angular cut (OMc) is an efficient method that uses
more than 75\% of the information contained in the data 
if $N_s \agt 200$. 
We argue that this loss of information is well worth the gain of
model independence, and continue to use this method for the simulation
of the actual angular distribution at a water Cherenkov detector.
It turns out that this method is much more efficient than the
one used in Ref.~\cite{sato}.

One may add gadolinium to Super-Kamiokande in order to
tag neutrons and therefore reduce the background due to 
inverse beta decay events.
We quantify the increase in the pointing accuracy as a function 
of the neutron tagging efficiency $\epsilon_{\rm tag}$.
It is found that the accuracy increases by more than a factor of
two with $\epsilon_{\rm tag}= 0.8$ and by nearly a factor of 3 for
$\epsilon_{\rm tag}=0.95$. 
For $\epsilon_{\rm tag}> 0.95$,
the oxygen events  act as the major background and that
saturates the advantage of increasing the tagging efficiency
beyond this value.

The efficiency of the OMc method improves with a smaller
background-to-signal ratio, which makes this method even 
more useful at high tagging efficiencies.
It is also observed that at higher $\epsilon_{\rm tag}$ this method
attains its maximum efficiency level for a much smaller 
number of events.
The optimal value of the cut increases with increasing 
$\epsilon_{\rm tag}$, though this dependence is weak and we perform
our estimations with a fixed value of the cut, which is near 
optimal in the whole parameter range. Our estimations are therefore
slightly conservative.

With a simulation of a water Cherenkov detector like Super-Kamiokande, 
we determine the pointing accuracy obtained from a SN at 10 kpc. 
The accuracy has a weak dependence on the neutrino mixing scenarios
and the initial neutrino spectra. We find that the worst case
scenario is when the detected $\nuebar$ spectrum has the largest
admixture of the initial $\nux$ spectrum, and the detected $\nue$
spectrum has the lowest admixture of the initial $\nux$ spectrum.
This worst case turns out to be the one with the inverted neutrino 
mass hierarchy, $\sin^2 \Theta_{13} \agt 10^{-3}$, and the largest 
possible solar mixing angle. 
The OMc method gives the pointing accuracy of  $7.8^\circ$ at
95\% C.L.\ without neutron tagging.
The accuracy improves to $3.6^\circ$ at 80\% tagging efficiency
and to $3.2^\circ$ at 95\% tagging efficiency.
Beyond this, the pointing accuracy saturates due to the presence 
of the oxygen events and the limited angular resolution of the detector.
For a larger detector, the expected accuracy may be scaled according
to the number of events. At a megaton detector like
Hyper-Kamiokande, this gives an accuracy of $0.6^\circ$ for
$\epsilon_{\rm tag} > 0.9$.

The SN shock wave may produce a TeV neutrino burst that arrives 
at the Earth within a day of the initial MeV neutrino signal.
This can give about 100 events with $E>1$ TeV at a km$^2$ detector
like IceCube. Since the angular resolution of this detector is
as good as 1$^\circ$, the SN may be located to an accuracy of 
a few tenths of a degree.
The limiting factor here is the alignment error of these detectors. 
The time correlation and the directionality of the events allows 
IceCube to detect them even ``from above''against
the background of atmospheric muons.


\begin{acknowledgments}

We thank Michael Altmann for suggesting the Landau distribution as
model of the angular resolution function of Super-Kamiokande and
Saunak Sen for clarifying some statistics issues.
We also thank Mathias Keil and Christian Spiering for useful
discussions and comments.
This work was
supported, in part, by the Deutsche Forschungsgemeinschaft under grant
No.\ SFB-375 and by the European Science Foundation (ESF) under the
Network Grant No.~86 Neutrino Astrophysics. M.K.\ and D.S.\
acknowledge support by an Emmy-Noether grant of the Deutsche
Forschungsgemeinschaft, R.T.\ a Marie-Curie-Fellowship of the European
Commission.

\end{acknowledgments}

\appendix
\section{Neutrino fluxes}
\label{nu}

For the time-integrated neutrino fluxes we assume distributions
of the form~\cite{Keil:2002in}
\begin{equation}\label{eq:spectralform}
F^0=
\frac{\Phi_0}{E_0}\,\frac{(1+\alpha)^{1+\alpha}}{\Gamma(1+\alpha)}  
\left(\frac{E}{E_0}\right)^\alpha 
\exp\left[-(\alpha+1)\frac{E}{E_0}\right] ~,
\end{equation}
where $F^0$ denotes the flux of a neutrino species emitted by the SN
scaled appropriately to the distance travelled from the SN to Earth.
Here $E_0$ is the average energy and $\alpha$ a parameter 
that relates to the width of the spectrum and
typically takes on values 2.5--5, depending on the flavor and the
phase of neutrino emission. The values of the total flux $\Phi_0$ and
the spectral parameters $\alpha$ and $E_0$ are generally different for
$\nu_e$, $\bar\nu_e$ and $\nu_x$, where $\nu_x$ stands for any of
$\nu_{\mu,\tau}$ or $\bar\nu_{\mu,\tau}$.

We consider two models for the initial neutrino fluxes. 
The first one is the recent calculation from the Garching group 
\cite{garching-model},
which we refer to as the model G. It includes all relevant
neutrino interaction rates, including nucleon bremsstahlung, 
neutrino pair processes, weak magnetism, nucleon recoils and
nuclear correlation effects. 
The second one is the Livermore simulation \cite{livermore-model},
referred to as model L,
that represents traditional predictions for flavor-dependent
SN neutrino spectra that have been used in many previous analyses.
The parameters of these models
are shown in Table~\ref{tab:models}. 
We take $\alpha(\nu_e)=\alpha(\bar{\nu}_e)= \alpha(\nu_x)=3.0$
for both models. The value of $\alpha$ is not expected to have
any significant influence on the pointing accuracy.

\begin{table}
\caption{The parameters in the neutrino spectra models from the
Garching group and the Livermore group.
\label{tab:models}}
\begin{ruledtabular}
\begin{tabular}{lccccc}
Model & $\langle E_0(\nu_e) \rangle$ & $\langle E_0(\nuebar) \rangle$&
$\langle E_0(\nux) \rangle$ & {\Large $\frac{\Phi_0(\nu_e)}{\Phi_0(\nu_x)}$} &
{\Large $\frac{\Phi_0(\nuebar)}{\Phi_0(\nu_x)}$}\\
\hline
G & 12 & 15 & 18 & 0.8 & 0.8 \\
L & 12 & 15 & 24 & 2.0 & 1.6 \\
\end{tabular}
\end{ruledtabular}
\end{table}

When neutrino mixing is taken into account, the fluxes arriving at 
a detector are
\begin{eqnarray}
F_{\nue} & = & p F_{\nue}^0 + (1-p) F_{\nux}^0 ~, \\ 
F_{\nuebar} & =  &\bar{p} F_{\nuebar}^0 + (1-\bar{p}) F_{\nu_x}^0 ~, \\
4 F_{\nux} & = & (1-p) F_{\nue}^0 + (1-\bar{p}) F_{\nuebar}^0 +
(2 + p + \bar{p}) F_{\nu_x}^0 ~. \quad \quad
\label{feDbar}
\end{eqnarray}
Since the four neutrino species $\nu_x$ cannot be distinguished at the
detectors, we only give the sum of their fluxes, $4 F_{\nux}$.
Here $p$ and $\bar{p}$ are the
survival probablilities of $\nue$  and $\nuebar$ respectively. 

Depending on the mass hierarchy and the value of the mixing angle
$\Theta_{13}$, the survival probabilities $p$ and $\bar{p}$ belong 
to one of the three mixing scenarios shown in Table~\ref{3comb}.
We neglect the Earth matter effects and the details of the
``transition'' region around $\sin^2 \Theta_{13} \sim 10^{-3}$
\cite{Dighe:1999bi,Lunardini:2001pb}. 
We also neglect terms of order $(\Theta_{13})^2$ in $p$ and $\bar{p}$.
Then $p$ and $\bar{p}$
depend only on the solar mixing angle as given in the table, 
and are independent of the values of solar and atmospheric mass 
squared differences as well as the atmospheric mixing angle.

\begin{table}
\caption{The possible combinations of survival probabilities
$p$ and $\bar{p}$.
\label{3comb}}
\begin{ruledtabular}
\begin{tabular}{llccc}
Case & Hierarchy &  $\sin^2 \Theta_{13}$  & $p$ &  $\bar{p}$ \\
\hline
(a) &  Normal & $\agt 10^{-3}$  & 0  & $\cos^2\Theta_\odot$ \\
(b) & Inverted &  $\agt 10^{-3}$ &  $\sin^2\Theta_\odot$ &  0 \\
(c) & Any & $\alt 10^{-3}$  & $\sin^2\Theta_\odot$ 
&  $\cos^2\Theta_\odot$ \\
\end{tabular}
\end{ruledtabular}
\end{table}

\section{Neutrino reactions in water}

\label{cross}

\subsection{Elastic Scattering on Electrons}

The differential cross-section of the reaction $\nu+e^-\to \nu+e^-$
with $\nu=\{\nu_e,\bar\nu_e,\nu_{\mu,\tau},\bar\nu_{\mu,\tau}\}$ is
given by
\be   \label{el}
 \frac{d\sigma}{dy} = 
 \frac{G_F^2 m_e E_\nu}{2\pi} \left[ A+B\,(1-y)^2-C\,\frac{m_e}{E_\nu}\,y
 \right] ~,
\ee         
where $y=E_e/E_\nu$ is the energy fraction transfered to the electron,
$G_F$ the Fermi constant, and $m_e$ the electron mass.  The
coefficients $A$, $B$ and $C$ differ for the four different reaction
channels and are given in Table~\ref{ABC}. The vector and axial-vector
coupling constants have the usual values $C_V
=-\frac{1}{2}+2\sin^2\Theta_W$ and $C_A=-\frac{1}{2}$ with
$\sin^2\Theta_W \approx 0.231$.

The scattering angle $\theta$ between the incoming neutrino and
final-state electron is implied by
\be
 y = \frac{2\,(m_e/E_\nu)\,\cos^2\theta}{(1+m_e/E_\nu)^2-\cos^2\theta}
 ~. 
\ee

\begin{table}
\caption{\label{ABC} 
Coefficients used in Eq.~(\ref{el}) for the elastic scattering of
neutrinos on electrons.}
\begin{ruledtabular}
\begin{tabular}{lccc}
   & $A$ & $B$ & $C$ \\[0.5ex] \hline  
$\nu_e$ & $(C_V{+}C_A{+}2)^2$ & $(C_V-C_A)^2$ & $(C_V+1)^2-(C_A+1)^2$ \\
$\bar\nu_e$ & $(C_V-C_A)^2$ & $(C_V{+}C_A{+}2)^2$ & $(C_V+1)^2-(C_A+1)^2$\\
$\nu_{\mu,\tau}$ & $(C_V+C_A)^2$ & $(C_V-C_A)^2$ & $C_V^2-C_A^2$\\
$\bar\nu_{\mu,\tau}$ & $(C_V-C_A)^2$ & $(C_V+C_A)^2$ & $C_V^2-C_A^2$ \\
\end{tabular}
\end{ruledtabular}
\end{table}

\subsection{Inverse Beta Decay}

For $\bar\nu_e p\to ne^+$ we use the differential and total
cross-sections Eqs.~(5--7) of Ref.~\cite{Strumia:2003zx},
including the leading QED radiative corrections.

\subsection{Oxygen as a Target}

Another important reaction is the charged-current $\nu_e$ absorption
on oxygen~\cite{Haxton:kc,Kolbe:2002gk}. The dominant channels are
\begin{eqnarray}
\nu_e+{}^{16}{\rm O}&\to&{}^{15}{\rm O}+p+e^-,\\
\nu_e+{}^{16}{\rm O}&\to&{}^{15}{\rm O}^*+p+\gamma+e^-,\\
\nu_e+{}^{16}{\rm O}&\to&{}^{14}{\rm N}^*+p+p+e^-\,.
\end{eqnarray}
While these reactions cause far fewer events in a water Cherenkov
detector than inverse beta decay, they do not have final-state
neutrons and thus cannot be tagged. Therefore, in a detector
configuration with efficient neutron tagging, these reactions provide
the dominant background to the directional electron scattering
reactions.

The neutrino energy threshold in these reactions is approximately
15~MeV.  The total cross section, summed over all channels, has been
tabulated for the range $15\leq E_\nu\leq 100$~MeV
\cite{Kolbe:2002gk}.  Directly above threshold the cross section is
very small.  We find that for $25\leq E_\nu\leq100$~MeV the tabulated
cross sections are nicely represented by the analytic fit
\begin{eqnarray}
&&\sigma\left(\nu_e+{}^{16}{\rm O}\to{\rm X}+e^-\right)={}\nonumber\\
&&\,4.7\times10^{-40}~{\rm cm}^2\,
\left[\left(\frac{E_\nu}{{\rm MeV}}\right)^{1/4}-15^{1/4}\right]^6 ~.
\end{eqnarray}

For an accurate determination of the detector response one needs the
differential distribution of final-state $e^-$ energies and angular
directions for a given incident $E_\nu$.  Ref.~\cite{Haxton:kc}
provides extensive plots of such distributions after folding them with
thermal $E_\nu$ distributions.  This information is too indirect for
our purposes. Therefore, we limit our investigation to a schematic
implementation of this process where we assume that in every reaction
the final-state energy is $E_{e}=E_\nu-15$~MeV. For the angular
distribution we assume
\begin{equation}
\frac{d\sigma}{d\cos\theta}
=1-\frac{1+(E_e/25~{\rm MeV})^4}{3+(E_e/25~{\rm MeV})^4}
\cos\theta\,,
\end{equation}
where $\theta$ is the angle between incident $\nu_e$ and final-state
$e^-$. This means that for small energies the angular distribution is
proportional to $1-\frac{1}{3}\cos\theta$ while for large energies it
is $1-\cos\theta$, i.e.\ it becomes more backward peaked for high
energies. Our schematic approach roughly mimics the behavior shown in
Ref.~\cite{Haxton:kc}. Since the oxygen cross section is very energy
dependent, the contribution of this reaction to the pointing accuracy
depends sensitively on the neutrino energy spectrum and is thus very
uncertain anyway.

Another potentially important class of charged-current reactions
is~\cite{Haxton:kc,Kolbe:2002gk}
\begin{equation}
\bar\nu_e+{}^{16}{\rm O}\to X+e^+ ~.
\end{equation}
However, the contribution to the detector signal is somewhat smaller
than caused by the above $\nu_e$ reactions.  Moreover, the $\bar\nu_e$
reactions typically involve final-state neutrons and thus are rejected
by neutron tagging. One exception is
\begin{equation}
\bar\nu_e+{}^{16}{\rm O}\to{}^{16}{\rm N}+e^+ ~,
\end{equation}
but its contribution is small. Therefore, we neglect this entire class
of reactions in our study.

Another class of reactions 
is the neutral-current excitation of oxygen~\cite{Langanke:1995he}
\begin{equation}
\nu+{}^{16}{\rm O}\to \nu+X+\gamma ~.
\end{equation}
Most of these reactions cannot be rejected by neutron tagging.
However, the total cross section for neutral-current scattering, 
including the channels without final-state $\gamma$, is smaller than 
for the charged-current $\nu_e$ reaction~\cite{Kolbe:2002gk}. 
Moreover, the $\gamma$ energies are below 10~MeV, and most of them 
even below our analysis threshold of 7~MeV. Therefore, we also neglect this
class of reactions.

\section{Event generation}
\label{ang-res}

For the generation of each of the events the following steps are performed: 
\begin{enumerate}
\item
The energy $E_\nu$ of the reacting neutrino is chosen according to
$F_\nu(E_\nu)\sigma(E_\nu)$. 
The energy $E_e$ and the scattering angle $\theta$ of the outgoing
electron/positron is chosen according to the differential
cross-section of the particular reaction.
\item
The measured energy $E_{\rm det}$ of the scattered
electron/positron is determined by adding Gaussian noise with variance 
$\sigma(E_e)=\sqrt{E_e E_0}$, where $E_0=0.22$~MeV. 
If $E_{\rm det}<E_{\rm th}=7$ MeV, 
the event is not used in the data analysis. 
\item
The measured position $(\phi,\theta) $ of the event is simulated
according the angular resolution function of Super-Kamiokande.  
\end{enumerate}

\begin{figure}[h]
\epsfig{file=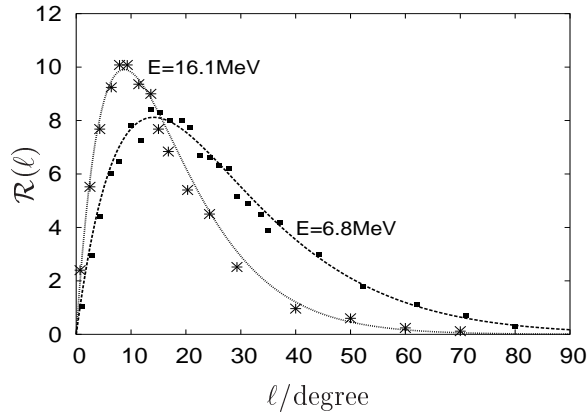,height=6.0cm}
\caption{
Our fit of the angular resolution ${\cal R}(\ell)$ as a function of
$\ell$ for positron energies $E_e=6.8$ and 16.1~MeV together
with measured points from Ref.~\cite{Nakahata:1998pz}.
\label{landau}}
\end{figure}

The angular resolution ${\cal R}(\ell)d\ell$ is defined as the
probability that inside two cones with opening angles $\ell$ and
$\ell+d\ell$ around the true direction the reconstructed direction is
contained. Reference~\cite{Nakahata:1998pz} gives numerical values for the
opening angle $\ell$ of a cone around the true direction which contains
68\% of the reconstructed directions as well as the values of 
${\cal R}(\ell)$ as a function of $\ell$ for various energies.
An inspection by eye shows that ${\cal R}(\ell)$ is characterized by a
large tail and cannot be well fitted by a Gaussian distribution.
Inspired by the Landau distribution for energy losses, we have found
that ${\cal R}(\ell)$ is well described by
\be
{\cal R}(\ell) = {\cal C}
 \exp\left(- \frac{x+ e^{-x}}{2} \right) \sin(\ell) ~,
\ee
where ${\cal C}$ is a normalization constant, and
\ba
 x &= & \frac{\ell}{\sigma}-a \,, \\
 a &=& -0.7E/{\rm MeV}-3.7 \,,\\
 \theta_{\max} &=& 38^\circ \, \sqrt{{\rm MeV}/E} ~,\\
 \sigma &=& [\theta_{\max}+\ln(2-\sqrt{3})]/a ~.
\ea
For illustration, we show in Fig.~\ref{landau} the angular resolution
${\cal R}(\ell)$ as a function of $\ell$ for positron energies
$E_e=6.8$ and 16.1~MeV as implemented in our simulation,
together with the measurements extracted from \cite{Nakahata:1998pz}.


\end{document}